\documentclass[twocolumn,pra,prc,nofootinbib]{revtex4-1}
\usepackage{amsmath}    
\usepackage{graphicx}   
\usepackage{verbatim}   
\usepackage{color}      
\usepackage{subfigure}  
\usepackage{hyperref}   
\usepackage{bm} 
\usepackage{multirow}
\usepackage{csquotes}
\usepackage{lipsum}

\newcommand{\classoption}[1]{\texttt{#1}}

\newcommand{\PbPb}{\ensuremath{\mbox{Pb--Pb}}}
\newcommand{\sqrtSnn}{\ensuremath{\sqrt{s_{\mathrm{NN}}}\text{ = 2.76 TeV}}}
\newcommand{\pt}{\ensuremath{p_{\mathrm{T}}}}
\newcommand{\fq}{\ensuremath{F_{\mathrm{q}}}}

\newcommand{\gev}{{GeV/\textit{c}}}
\newcommand{\fqtwo}{\ensuremath{F_{2}}}

\expandafter\ifx\csname package@font\endcsname\relax\else
\expandafter\expandafter
\expandafter\usepackage
\expandafter\expandafter
\expandafter{\csname package@font\endcsname}%
\fi
\DeclareRobustCommand\substyle{\name@idx{document substyle}}%
\DeclareRobustCommand\classoption{\name@idx{document class option}}%
\DeclareRobustCommand\classname{\name@idx{document class}}%
\def\name@idx#1#2{
	{\ttfamily#2}%
	\index{#2\space#1=\string\ttt{#2}\space#1}\index{#1>#2=\string\ttt{#2}}%
}
\begin{document}
	
		\widetext
	\title{Intermittency study of charged particles generated in \PbPb~collisions at \sqrtSnn~using EPOS3}
	\author{Ramni Gupta and Salman Khurshid Malik}
	\email[mailto:]{maroozmalik@gmail.com} 
	\affiliation{Department of Physics, University of Jammu, Jammu (J\&K), India}
	\date{\today}
\begin{abstract}
	Charged particle multiplicity fluctuations in Pb-Pb collisions are studied for the central events generated using EPOS3 (hydro and hydro+cascade) at $\sqrtSnn$. Intermittency analysis is performed in the mid-rapidity region in two-dimensional ($\eta$, $\phi$) phase space within the narrow transverse momentum (\pt) bins in the low \pt~region (\pt~$\leq~1.0~\gev$). Power-law scaling of the normalized factorial moments with the number of bins is not observed to be significant in any of the \pt~bin. Scaling exponent $\nu$, deduced for a few $\pt$ bins is greater than that of the value 1.304, predicted for the second order phase-transition by the  Ginzburg-Landau theory. The link in the notions of fractality is also studied. Fractal dimensions, $D_{q}$ are observed to decrease with the order of the moment $q$ suggesting the multifractal nature of the particle generation in EPOS3.
\end{abstract} 
	\maketitle
	\section{Introduction}
	\label{introduction}
	The strongly interacting dense state of matter, believed to represent QGP (quark-gluon plasma) after its creation in heavy-ion collision rapidly cools into a spray of particles. This array of particles carry signals of QGP, and its properties which can be directly and indirectly measured by detectors that are encircling the collision point. Of the myriad of analysis tools to understand the dynamics of this particle production~\cite{Sarkar:2010zza} and phase changes in the matter while passing into the QGP phase from the hadronic phase and vice versa, an important one is the fluctuations study of the observables. Lattice QCD predicts large fluctuations being associated with the system undergoing phase transition. Multiplicity distributions characterize the system formed or any phase transition in the heavy ion collisions. 
	Studies of multiplicity fluctuations have prompted considerable advances in this area of research. Large particle density fluctuations in the JACEE event~\cite{Burnett:1983pb} and its explanation by normalized factorial moments triggered investigations of multiplicity fluctuation patterns in multihadronic events with decreasing domains of phase-space~\cite{Kittel:2005fu}. The presence of power-law behaviour or scale-invariance of normalized factorial moments with decreasing phase-space interval or increasing bins is termed as \emph{intermittency}~\citep{Bialas:1985jb,Bialas:1988wc}. Observation of intermittency signals the presence of self-similar, fractal nature of the particle production. If fluctuations have a dynamical origin, the underlying probability density will be reflected as intermittency behaviour. The existence of dynamical fluctuations can thus be studied using normalized factorial moments (NFMs)~\cite{Bialas:1985jb} in one, two or three-dimensional phase-space. 
	\par 
	The idea of intermittency has been obtained from the theory of turbulent flow. There it signifies as a property of turbulent fluid: vortices of fluid with different size alternate in such a way that they form  self-similar structures. These vortices do not necessarily fill in the entire volume, but they instead create an intermittent pattern in the regions of laminar flow. This property is given by a power-law variation of the vortex-distribution moments on their size. So, the self-similar nature of vortices directly creates a relation between intermittency and fractality. Self-similar objects of non-integral dimensions are called \emph{fractals}~\cite{fractals}. A fractal dimension is a generalization of an ordinary topological dimensionality to non-integers.
	\par 
	The proposal to look for intermittency also prompts a thorough study of phase-transition models. 
	A very straightforward model that offers some hint on the nature of a second-order phase transition is the Ising model in two dimensions~\cite{Lindaereichl}. Intermittency in Ising model has been studied both analytically and numerically~\citep{Satz:1989vj,Bambah:1989fy} and the anomalous fractal dimension ($d_q$) is found to be 1/8, independent of the order of moment, $q$. It has been conjectured on this acount that intermittency may be monofractal in QCD second order phase transition~\cite{Bialas:1990xd}. However, all types of interactions including heavy-ion collisions show multifractal behaviour~\citep{Kittel:2005fu,DeWolf:1995nyp}. Also, Yang-Mills fields have been applied to QCD within asymptotic approximation where the fractal dimension is determined as a function of entropic index and value obtained for entropic index is in good agreement with the experimental data~\cite{Deppman:2019klo}. For first order phase transition, all $d_q$ are zero and no intermittency was observed. Intermittency has also been studied in Ginzburg-Landau (GL) theory, which has been accustomed to explain the confinement of magnetic fields into fluxoids in a type-II superconductor. From the study of normalized factorial moments with decreasing phase space bins for the Ginzburg-Landau second order formalism, the anomalous fractal dimension is observed not to be constant. It follows $d_q$/$d_2$ = $(q-1)^{(\nu-1)}$, where $\nu$ is the scaling exponent~\cite{Hwa:1992uq}. $\nu$ is observed to be a universal quantity valid for all systems describable by the GL theory for second order phase transition and it is independent of the underlying dimensions or the parameters of the model.  This is of particular importance for a QCD phase transition, since neither the transition temperature nor the other important parameters are known there. If a signature of quark-hadron phase transition depends on the details of the heavy-ion collisions e.g. nuclear sizes, collision energy, transverse energy, etc., then even after the system has passed the thresholds for the creation of QGP, such a signature is likely to be sensitive to this theory. 
	
	\par 
	In this work, intermittency analysis is performed for the charged particles generated  in the midrapidity region of the central events (b~$\leq$~3.5~$fm$) from \PbPb~collisions at \sqrtSnn~using EPOS3 (hydro) and EPOS3 (hydro+cascade).
	\par 
	The plan of the paper is; EPOS3 model~\citep{Werner:2013tya} is introduced in Section \ref{epos3}. The methodology of  analysis  is given in Section \ref{methodolgy}. In Section \ref{discussion} observations and results are given followed by a summary in Section \ref{summ}.

	\section{A brief introduction to EPOS3}
	\label{epos3}
	
	EPOS3~\citep{Werner:2010aa,Werner:2013tya,Werner:2007bf} is a hybrid Monte-Carlo event generator with a 3$+$1D hydrodynamical expanding system. This model is based on flux tube initial conditions which are generated in the Gribov-Regge multiple scattering framework. The formalism is referred to as \enquote{Parton based Gribov Regge Theory}, which is detailed in~\citep{Drescher:2000ha}. An individual scattering gives rise to a parton-ladder and is called a \textit{Pomeron}. Each parton ladder eventually shows up as flux tubes (or strings) and is identified by a pQCD hard process, plus initial and final state linear parton emission. Saturation scale, $Q_s$ is employed to consider non-linear effects. This depends upon the energy and the number of participants attached to the pomeron under consideration.  
	\par 
	For a pomeron, after multiple scatterings the final state partonic system has two colour flux tubes, mainly longitudinal with transversely moving pieces carrying transverse momentum of the hard scattered partons. Each pomeron by virtue of its cylindrical topology has two flux tubes. The flux tubes also expand with time and gets fragmented into string segments of quark-antiquark pairs, resulting in more than two flux tubes. The high string-density areas form the \enquote{\textit{core}}(bulk matter)~\citep{Werner:2007bf} and the low string-density areas form the \enquote{\textit{corona}}. The corona particles originate from the string decay by Schwinger mechanism. In EPOS3, only the core region thermalizes, flows and hadronizes. The core undergoes viscous hydrodynamic evolution and as the hadronisation temperature ($T_H =$168 MeV) is reached, Cooper-Frye mechanism~\citep{Cooper:1974mv} is applied to convert the fluid into particles. For hadronic cascade, all the hadrons participate from both core and corona. When the cascading mechanism is included in the modeling, EPOS3 might show self-similarity and thus intermittency effect~\cite{eneziano}. EPOS3 is universal and unique in the sense that it treats \textit{pp}, \textit{pA} and \textit{AA} scatterings with the same core-corona procedure. 
	\par
\begin{figure}[h!]
		\centering
		{
			\includegraphics[scale=0.45]{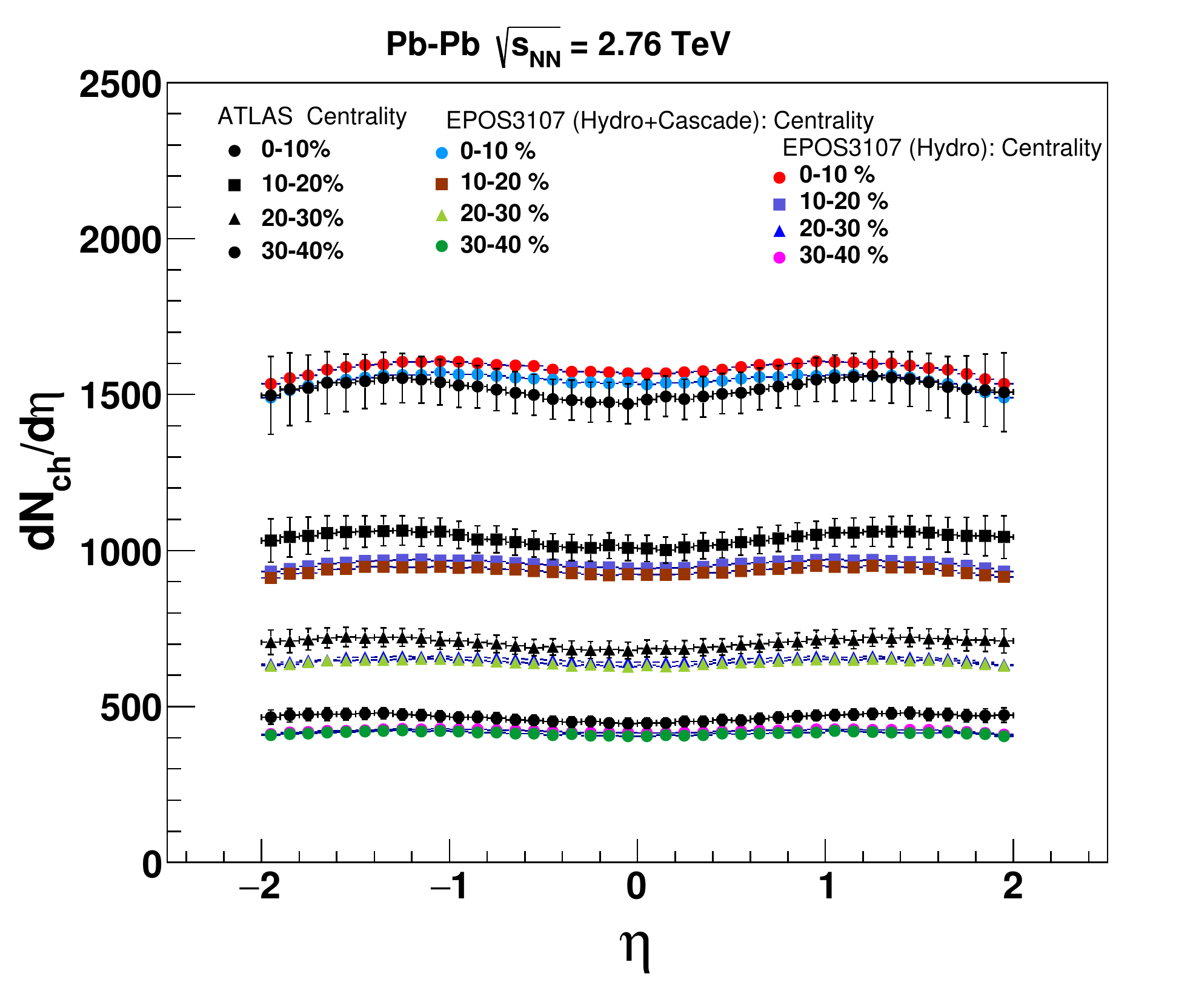}}
		
		\caption{\label{fig:atlas}Charged particle pseudorapidity density distributions of EPOS3 (hydro)  and EPOS3 (hydro+cascade) compared with that of the ATLAS data~\cite{ATLAS:2011ag}, for \PbPb~collisions at \sqrtSnn.}
	\end{figure}

	A sample of 66,350 and 23,502 minimum biased events have been generated for \PbPb~collisions at \sqrtSnn~using the hydro and the hydro+cascade mode of the EPOS3. The charged particle pseudorapidity density ($dN_{ch}/d\eta$) distributions of these events are shown in Figure \ref{fig:atlas}, for various centralities and are compared with that of ATLAS data~\cite{ATLAS:2011ag} for the same system and energy. Where for the polar angle $\theta$ of the particle, measured with respect to the beam axis, the pseudorapidity ($\eta$) is defined as $\eta = -\ln tan (\frac{\theta}{2})$. In this work, analysis is performed for the charged particles generated in full azimuthal space with $|\eta| \le 0.8|$ in the most central events. It is observed (Figure \ref{fig:atlas}) that in the midrapidity region of our interest ($|\eta|\leq 0.8$), charged particle pseudorapidity density of the EPOS3 generated central (0-10\%) events, slightly overestimates the ATLAS data within errors. 
\par
Intermittency studies at low energies, had limitation of statistics because a lesser number of particles were available per bin for the order of the moment ~$q$~$\geq$~2. In the present collider experiments, with the availability of high multiplicity events per pseudorapidity unit both in $pp$ and $AA$ collisions the studies of local multiplicity fluctuations, dependent on the bin contents can be taken up, to get a clear and complete picture of the multiparticle production. 
Predictions for intermittency analysis of data at present collider energies are still not available. Present work is carried to study scaling behaviours of the charged particles multiplicity fluctuations and  hence the intermittency in the EPOS3 model, which is based on the hydrodynamic particle production mechanism.
	 
		\section{Methodology}
	\label{methodolgy}
	Observation of spike events first noticed in the cosmic ray interaction~\citep{Burnett:1983pb} and later in the laboratory~\citep{Kittel:2005fu,DeWolf:1995nyp} lead to great spurt of interest in the studies of intermittency in particle production in high-energy collisions. In~\citep{Bialas:1988wc,Bialas:1985jb}, groundbreaking work was done theoretically formulating the features of intermittency in the field of particle physics. 
	\par 
Intermittency is defined as the scale-invariance of NFM, \fq,~with respect to changes in the size of phase-space cells (bins)~\citep{Bialas:1985jb}. For one-dimensional phase-space of rapidity $Y$, with cell $\delta y$ (say), it is defined as:
	\begin{eqnarray}
		F_q(\delta y) \,\propto\, (\delta y)^{-\phi_{q}} \hspace{5mm} (\delta y\rightarrow 0).
		\label{eqn:intermittency}
	\end{eqnarray}
	where $F_q$'s are the NFM~\citep{Bialas:1985jb}, of order $q$, where $q$ is a positive integer and takes values $\geq 2$ and $\phi_{q}>0$ is called the \enquote{intermittency index} or \enquote{intermittency slope}. In terms of the number of bins $M$ in the phase space, where $M \propto 1/\delta$, Equation (\ref{eqn:intermittency}) can be written as:
	\begin{eqnarray}
		F_q(M) \,\propto\, M^{\phi_{q}}.
		\label{eqn:Mscaling}
	\end{eqnarray}
	 In~\citep{Hwa:2016khr,Hwa:2014jea} it is proposed that NFM using \emph{event} NFM be investigated at LHC energies where the charged particle density is very high. The event NFM, $F_q^e$ are defined as: 
	\begin{eqnarray}
		F_q^e(M)\,=\, \frac{f_q^e(M)}{\Big[f_1^e(M)\Big]^q},
		\label{eqn:37}
	\end{eqnarray}
	with $f_q^e(M)=\left<n_m(n_m-1)\dots\dots(n_m-q+1)\right>_e$, where $\left< \dots \right>_e$ is the averaging over all bins in an e${^{th}}$ event, called horizontal averaging and $n_m$ is bin multiplicity of the m$^{th}$ bin. NFM $F_q$ for a sample of events, $N_{evt}$ is then:
	\begin{eqnarray}
		F_q(M)\,=\, \frac{1}{N_{evt}}\sum_{e=1}^{N_{evt}}F_q^e(M).
		\label{eqn:38}
	\end{eqnarray}
	$F_q(M)$ enjoys the property of filtering out statistical fluctuations (or noise)~\citep{Bialas:1985jb,Hwa:2011bu}. The scaling of the NFM, $F_q$, with number of bins $M$ as in Equation (\ref{eqn:Mscaling}) is referred here as \emph{M-scaling}. Observation of this scaling would indicate the self-similarity in the spatial distribution of the particles. It has been observed that the Ginzburg-Landau formalism~\citep{Hwa:1992uq} for second-order phase transition, $F_q$ follows power-law as:
	\par 
	\begin{align}
		F_q \propto F_{2}^{\beta_q},
		\label{eqn:fscaling}
	\end{align}
	such that $\beta_q= (q-1)^\nu~\text{with}~\nu=1.304$.
	Equation (\ref{eqn:fscaling}) is referred here as \emph{F-scaling}. Its validity is independent of the scaling behaviour in Equation (\ref{eqn:Mscaling}).
	\par 
	There exist more complicated self-similar objects which include fractal patterns with different non-integer dimensions, \emph{multifractals}~\citep{PALADIN1987147,Feder1988,Kittel:2005fu,DeWolf:1995nyp}. Multifractals are characterized by generalized (or R$\grave{e}$nyi) dimensions ($D_q$) which are decreasing functions of $q$. The thought of R$\grave{e}$nyi dimensions $D_q$ generalizes the idea of fractal dimension $D_0=D_F$, information dimension $D_1$ and correlation dimension $D_2$. Consequently, the R$\grave{e}$nyi dimension is often known as the generalized dimension. The anomalous fractal dimension ($d_q$) is related to the generalized dimension ($D_q$) by the relation:
	 
	\begin{align} d_q \,=\, D-D_q
	\end{align} 
	\par
	where, $D$ is the topological dimension that represents the number of dimensions. A relation between the exponents of factorial moments, intermittency index ($\phi_{q}$) and generalized moments can be devised at low values of $q$ as: 
	\par
	\begin{align}\phi_q + \tau(q) \,=\, (q-1)D
	\end{align}
	where the exponents are related to R$\grave{e}$nyi dimensions and codimension as:\par
	\begin{align}
		\tau(q) \,=\, (q-1)D_q \hspace{3mm}
		\text{and} \hspace{3mm} \phi_{q} \,=\, (q-1)d_q
	\end{align} It is needed to stress that the slope $\tau_{q}$ has no dynamical feature of $\phi_{q}$ and needs to be corrected for the statistical contribution to be removed~\cite{Chiu:1990bc}. Increasing $d_q$ with $q$ is a signal of the multifractal system. 
	\par 
	Here, intermittency and notion of fractality for charged particle multiplicity distribution is studied in the two-dimensional phase space ($\eta, \phi$) of the events generated using EPOS3 for the \PbPb~collision system at \sqrtSnn.
	\section{Analysis and Observations}
	\label{discussion}
A two dimensional intermittency analysis in ($\eta, \phi$) phase space in different~\pt~\footnote{$p_{T} = \sqrt{p_{x}^{2}+p_{y}^{2}}$, where $p_{x}$ and $p_{y}$ are the momentum components in the transverse momentum plane} bins of varying widths (0.2 $\leq \pt \leq$ 0.4 \gev, 0.4 $\leq \pt \leq$ 0.6 \gev, 0.6 $\leq \pt \leq$ 0.8 \gev, 0.8 $\leq \pt \leq$ 1.0 \gev, 0.2 $\leq \pt \leq$ 0.6 \gev, 0.2 $\leq \pt \leq$ 0.8 \gev~and 0.2 $\leq \pt \leq$ 1.0 \gev) is performed for two event samples for \PbPb~collisions at \sqrtSnn~generated using two modes of EPOS3. Central events with impact parameter $b~\leq~$3.5 $fm$ have been analyzed. In this work, charged particles (pions, kaons, protons) generated in the kinematical region with $|\eta|\leq 0.8$, full $\phi$ coverage and $\pt \leq$ 1.0 \gev~have been studied.
	\begin{figure}[h!]
		\centering
		\subfigure[\ 0.4 $\leq \pt \leq$ 0.6 \gev.]{
			\includegraphics[scale=0.2]{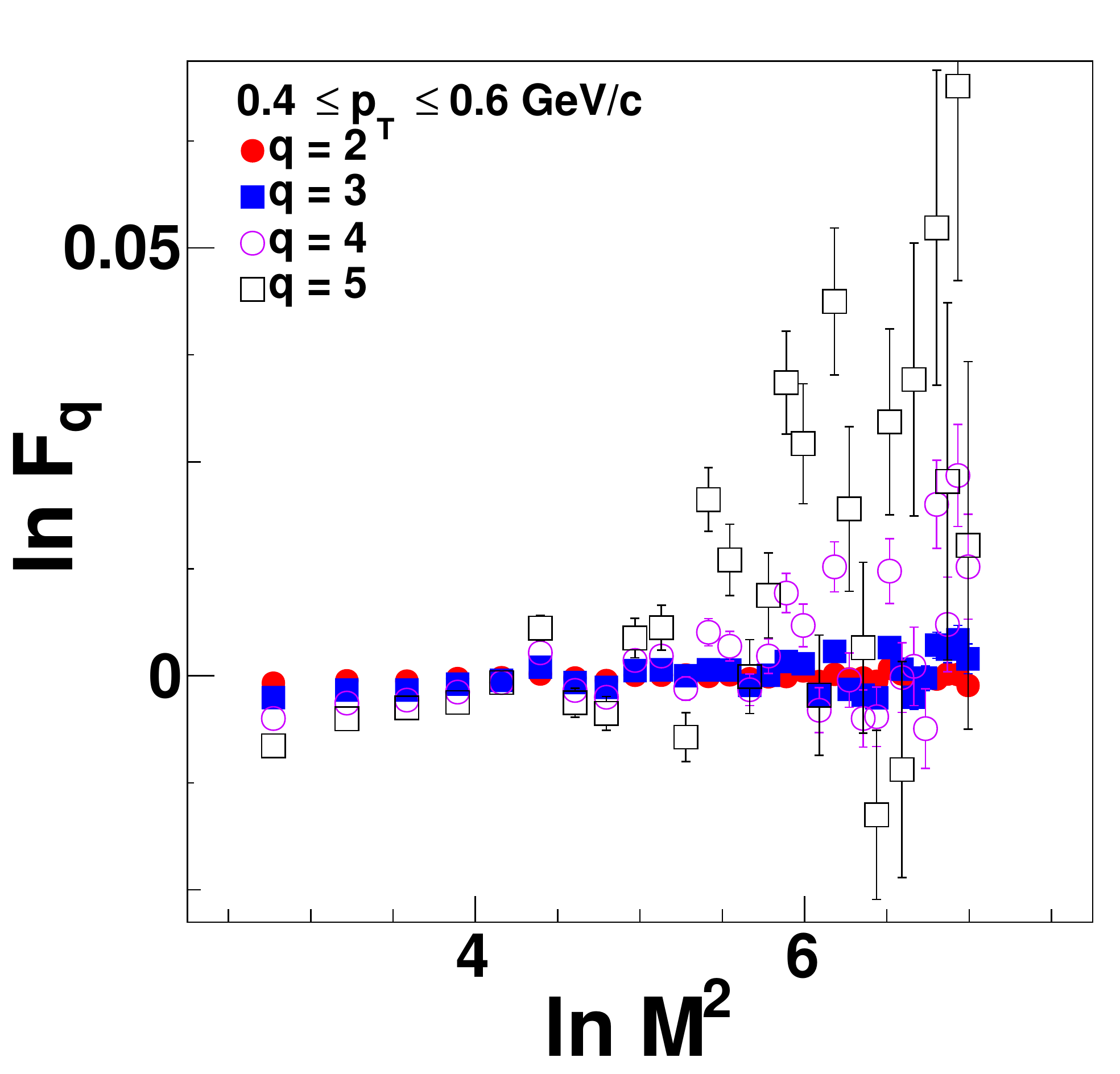}}
		\subfigure[\ 0.6 $\leq \pt \leq$ 0.8 \gev.]{
			\includegraphics[scale=0.2]{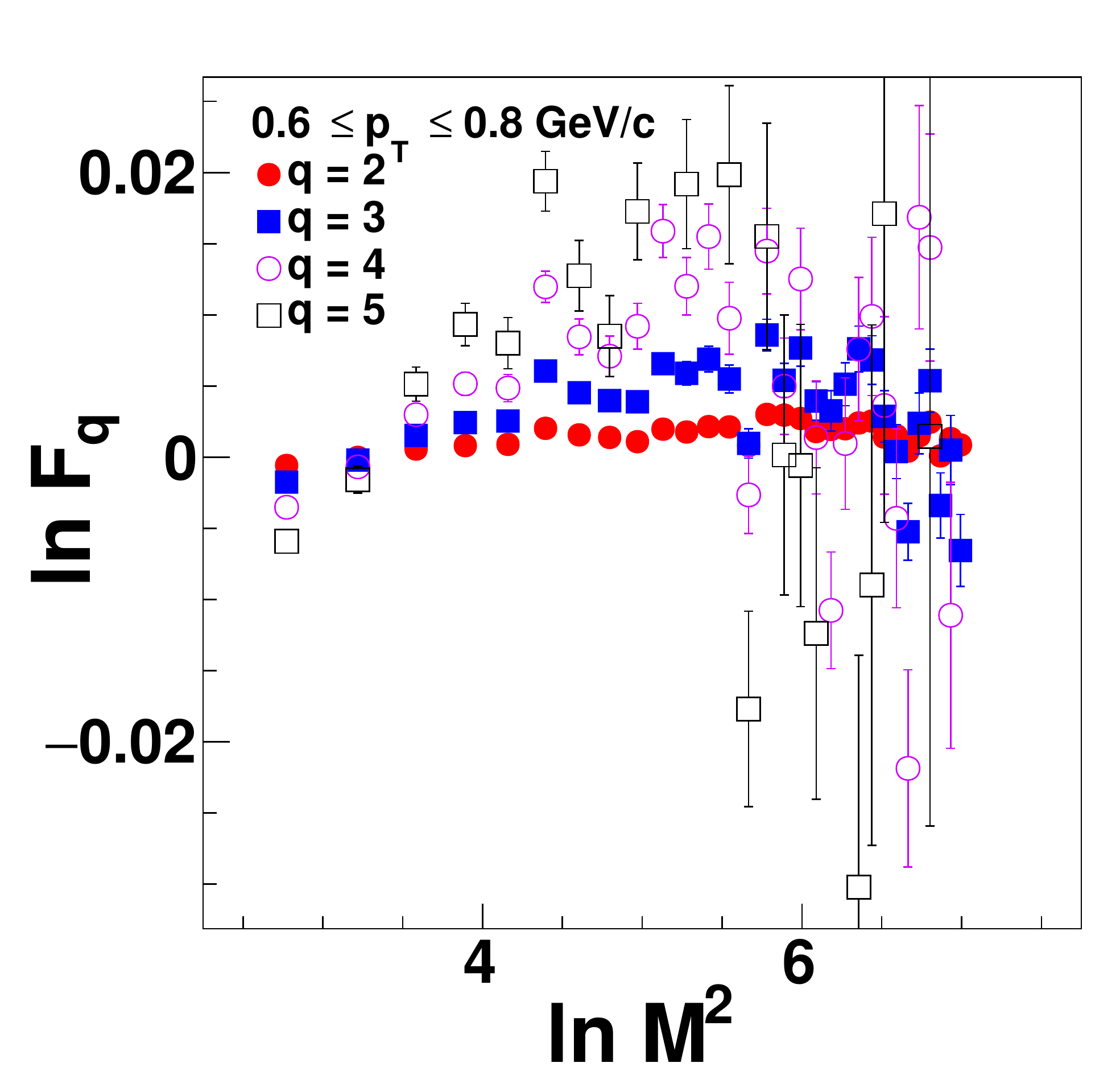}}
		\caption{\label{fig:ms1/hydro} Log-Log~\fq~dependence on number of bins ($M^2$) for EPOS3-hydro events for the \pt~bins 0.4 $\leq \pt \leq$ 0.6 \gev~and 0.6 $\leq \pt \leq$ 0.8 \gev.}
	\end{figure}
	\begin{figure}[h!]
	\centering
	\subfigure[\ 0.4 $\leq \pt \leq$ 0.6 \gev.]{
		\includegraphics[scale=0.2]{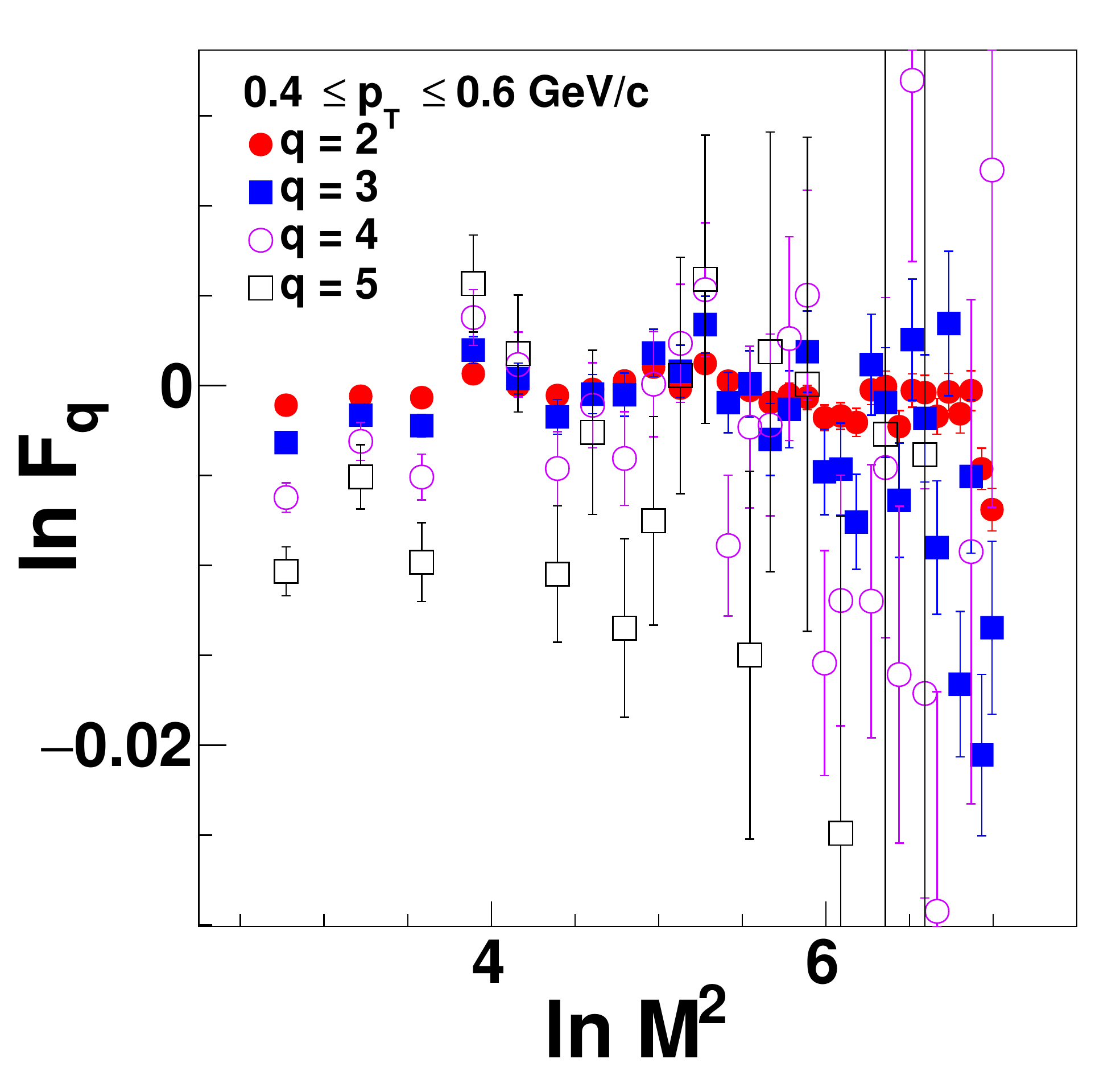}}
	\subfigure[\ 0.6 $\leq \pt \leq$ 0.8 \gev.]{
		\includegraphics[scale=0.2]{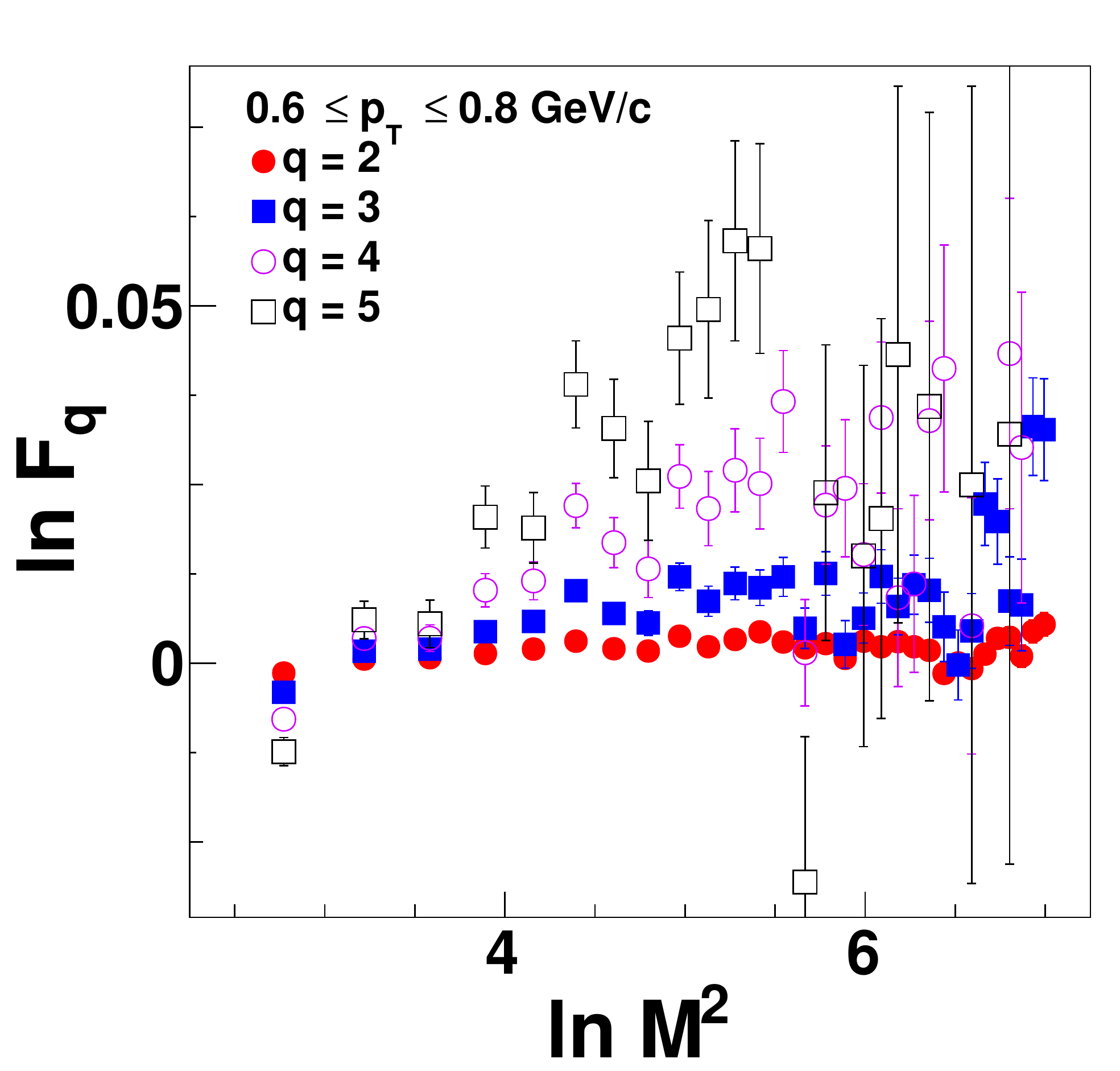}}
	\caption{\label{fig:ms1/hydrocascade} Log-Log~\fq~dependence on number of bins ($M^2$) for EPOS3-hydro+cascade events for the \pt~bins 0.4 $\leq \pt \leq$ 0.6 \gev~and 0.6 $\leq \pt \leq$ 0.8 \gev.}
	\end{figure}

	The methodology adopted for analysis is same as in~\cite{Sharma:2018vtf} for the SM AMPT model. The ($\eta$,$\phi$) phase space in a p$_T$ bin, for an event, is divided into a $M \times M$ matrix such that there are a total of $M^2$ bins. $M$ is taken from 2 to 32 in an interval of 2. Number of charged particles in a bin, $n_m$, is the bin multiplicity in the $m^{th}$ bin. Event factorial moment, $F_q^e(M)$ (Equation (\ref{eqn:37})) is determined for $n_m \geq q$, where $q= 2, 3, 4,~\text{and}~5$ is the order of the moment. $F_q^e(M)$ are obtained for all the events in the event sample. This gives the event factorial moment distribution and hence the $\fq(M)$ (Equation (\ref{eqn:38})). $\fq(M)$'s are thus studied for their dependence on $M$ and the second order normalized factorial moments (\fqtwo(M)).
	\begin{figure}[h!]
		\centering
		\subfigure[\ 0.2 $\leq \pt \leq$ 0.8 \gev.]{
			\includegraphics[scale=0.2]{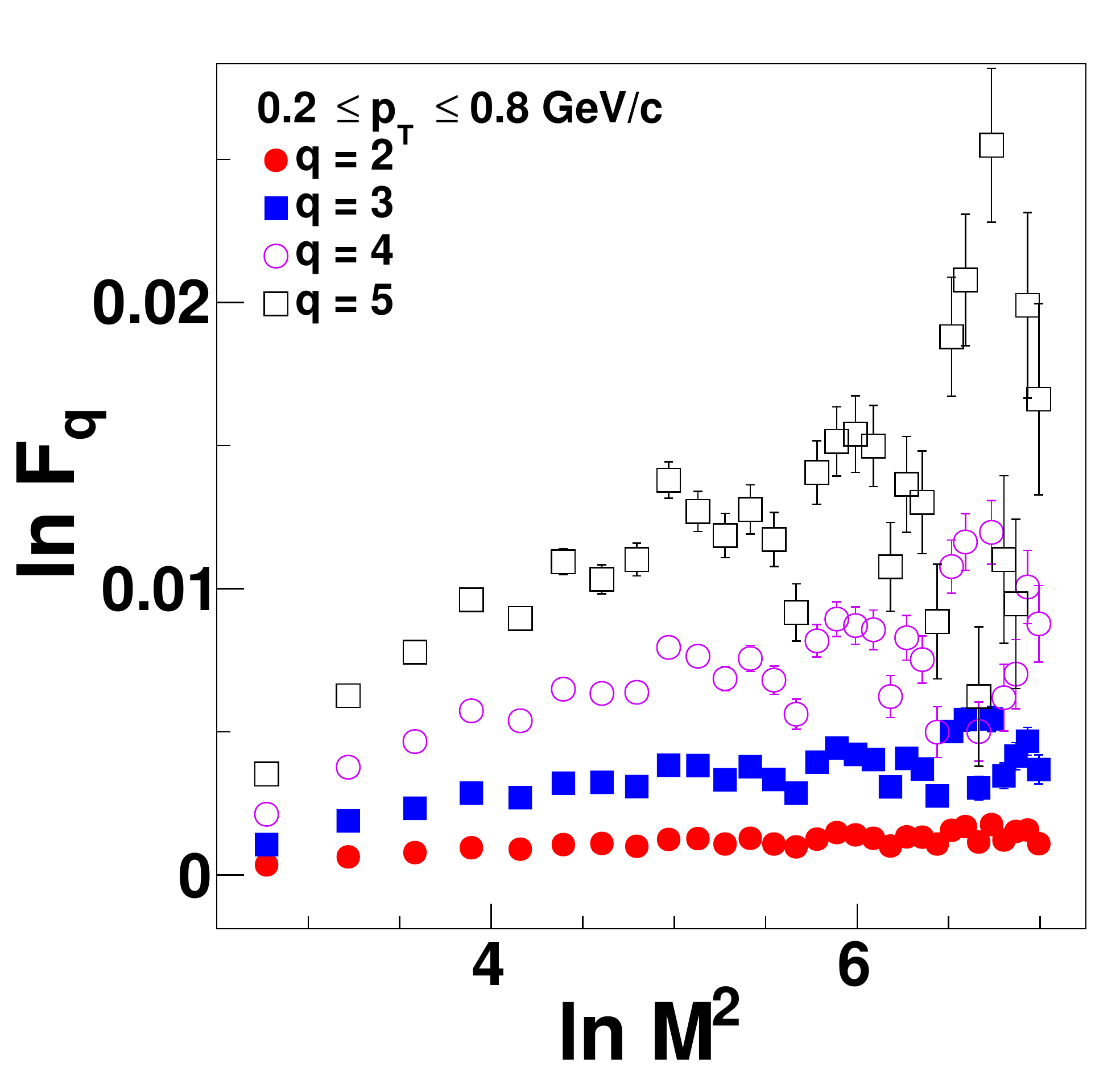}}
		\subfigure[\ 0.2 $\leq \pt \leq$ 1.0 \gev.]{
			\includegraphics[scale=0.2]{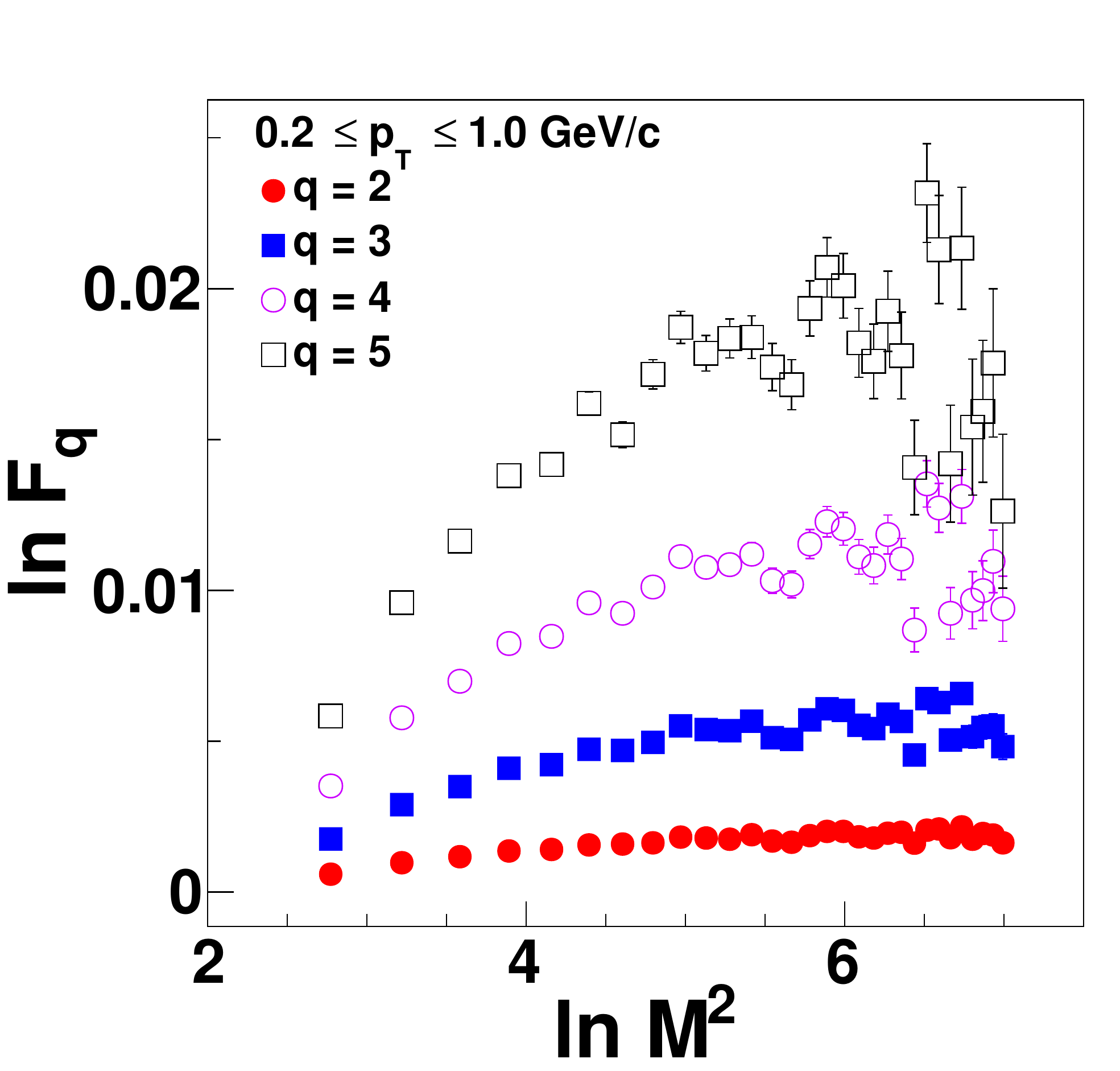}}
		\caption{\label{fig:ms/hydro} Log-Log~\fq~dependence on number of bins ($M^2$) for EPOS3-hydro events for the \pt~bins 0.2 $\leq \pt \leq$ 0.8 \gev~and 0.2 $\leq \pt \leq$ 1.0 \gev.}
	\end{figure}
\begin{figure}[h!]
	\centering
	\subfigure[\ 0.2 $\leq \pt \leq$ 0.8 \gev.]{
		\includegraphics[scale=0.2]{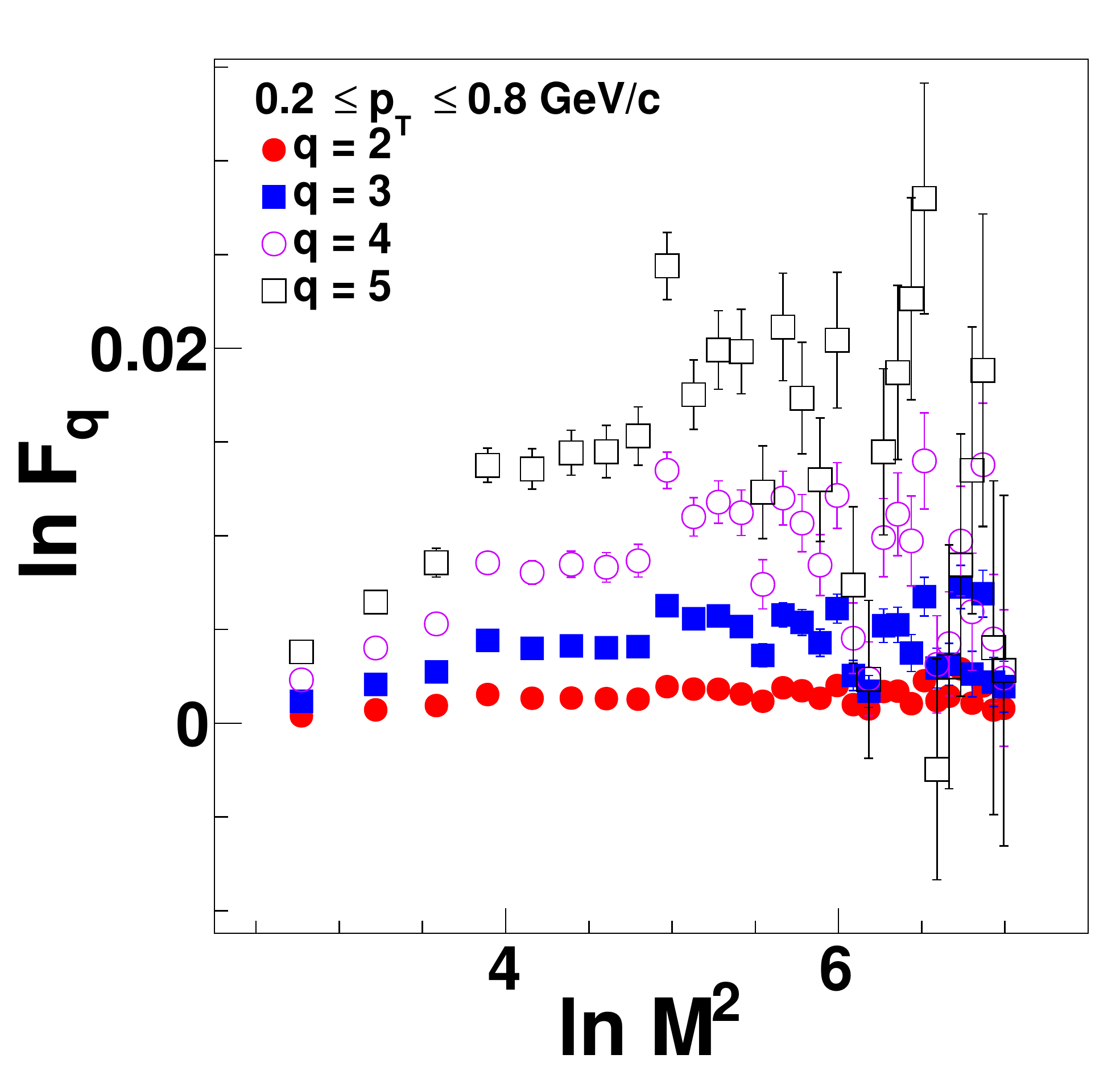}}
	\subfigure[\ 0.2 $\leq \pt \leq$ 1.0 \gev.]{
		\includegraphics[scale=0.2]{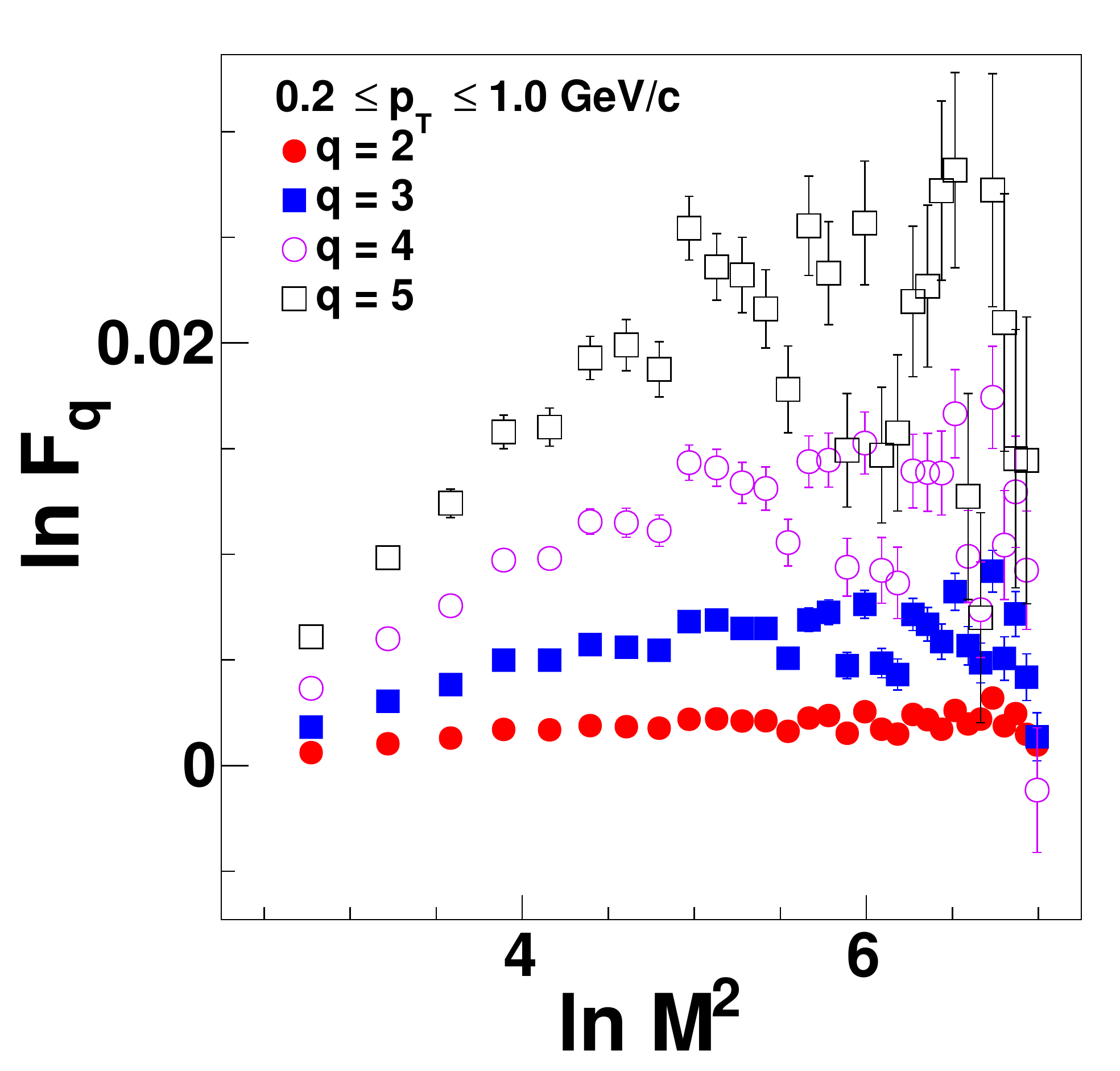}}
	\caption{\label{fig:ms/hydrocascade} Log-Log~\fq~dependence on number of bins ($M^2$) for EPOS3-hydro+cascade events for the \pt~bins 0.2 $\leq \pt \leq$ 0.8 \gev~and 0.2 $\leq \pt \leq$ 1.0 \gev.}
\end{figure}
\par
From the study of dependence of $F_q$ on $M$ (M-scaling) for the various \pt~bins, it is observed that for the small \pt~bins with width $\Delta\pt$=0.2 \gev~(0.2 $\leq \pt \leq$ 0.4 \gev, 0.4 $\leq \pt \leq$ 0.6 \gev, 0.6 $\leq \pt \leq$ 0.8 \gev~and 0.8 $\leq \pt \leq$ 1.0 \gev), M-scaling is absent in case of both hydro and hydro+cascade events. For two bins, $\ln F_{q}$ vs $\ln M^2$ graphs for $q=$2, 3, 4, 5 are given in Figure \ref{fig:ms1/hydro} (EPOS3 Hydro) and Figure \ref{fig:ms1/hydrocascade} (EPOS3 hydro+cascade). For the wider \pt~bins with $\Delta\pt \geq$0.6 \gev~that is for $0.2 \leq p_{T} \leq 0.8$ \gev~and $0.2\leq p_{T} \leq 1.0$ \gev, scaling of $F_q$ with M is observed in the lower $M$ region followed by saturation effects at higher $M$ region as observed in  $\ln \fq$ vs $\ln M^2$ graph in Figure \ref{fig:ms/hydro} for EPOS3(hydro). For the same \pt~bins that is  0.2$\leq \pt \leq$0.8 \gev~and 0.2$\leq \pt \leq$1.0 \gev~ Figure \ref{fig:ms/hydrocascade} shows the same graphs from EPOS3(hydro+cascade) events. M-scaling is observed to be present in the low $M$ region with saturation and overlapping effects at higher $M$. Absence of power-law or M-scaling in narrow~\pt~bins clearly indicates the absence of local density fluctuations and hence, the intermittency signal. The presence of weak intermittency in the wider~\pt~bins is probably due to number effect as average bin content increases in the given phase space. The error bars are the statistical uncertainties, calculated using the error propagation formula as suggested in~\citep{Metzger}.
\par
\begin{figure}[h!]
		\centering
		\subfigure[\ 0.2 $\leq \pt \leq$ 0.8 \gev.]{
			\includegraphics*[scale=0.2]{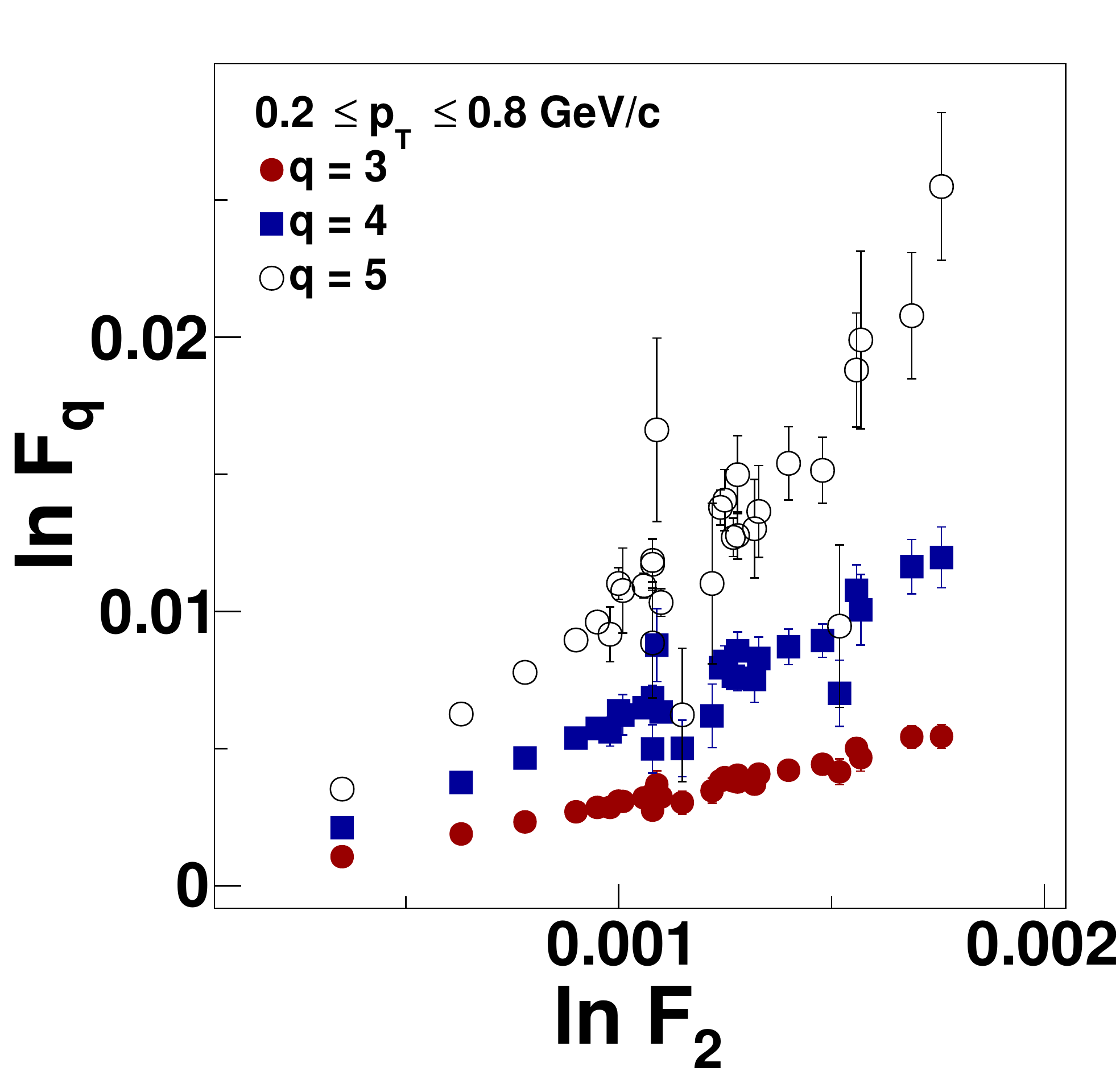}}
		\subfigure[\ 0.2 $\leq \pt \leq$ 1.0 \gev.]{
			\includegraphics*[scale=0.2 ]{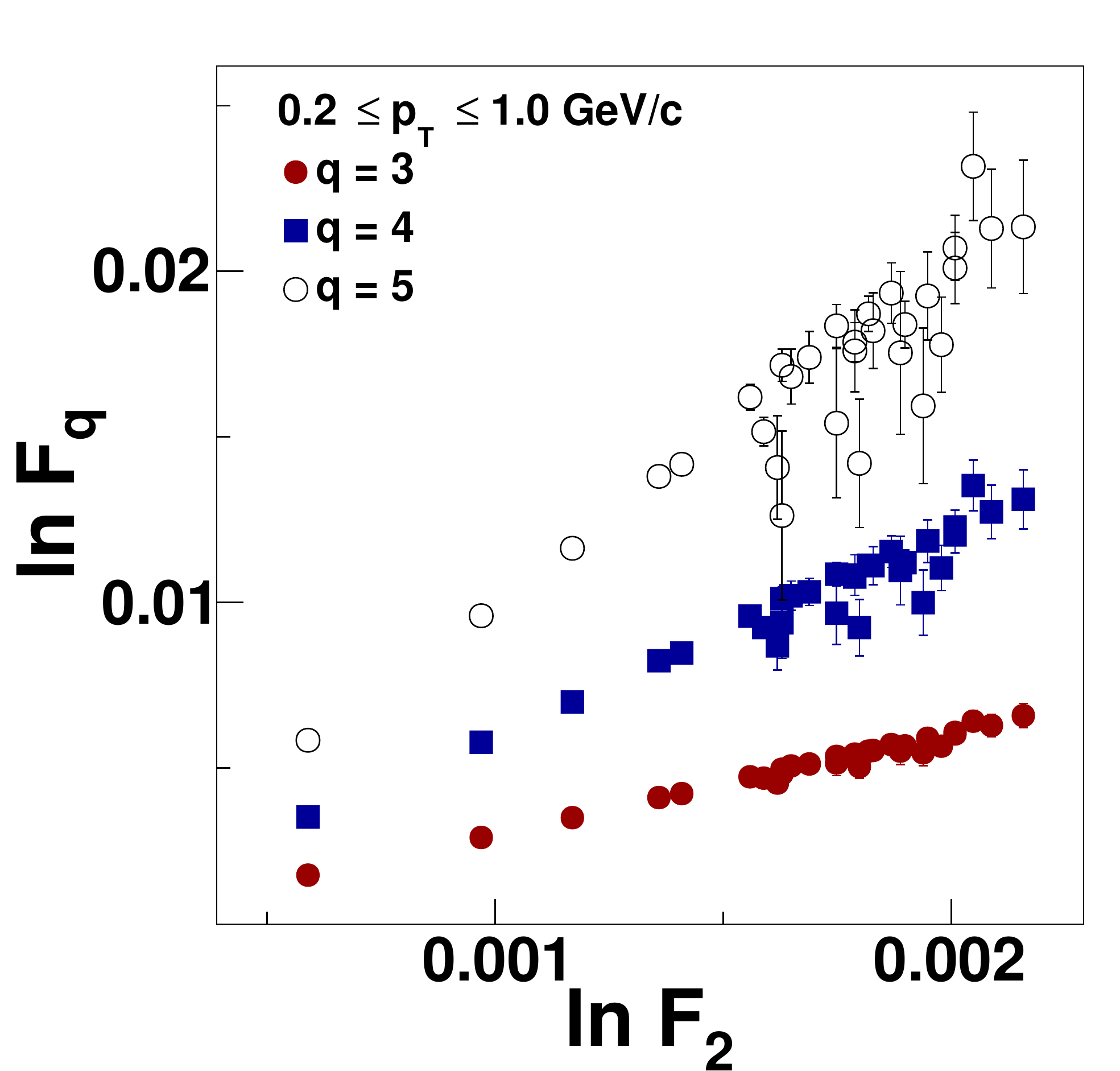}}
		\caption{\label{fig:fs/hydro} Log-Log~\fq~dependence on \fqtwo~for EPOS3-hydro events for the \pt~bins 0.2 $\leq \pt \leq$ 0.8 \gev~and 0.2 $\leq \pt \leq$ 1.0 \gev.}
	\end{figure}

\begin{figure}[h!]
		\centering
		\subfigure[\ 0.2 $\leq \pt \leq$ 0.8 \gev.]{
			\includegraphics*[scale=0.2]{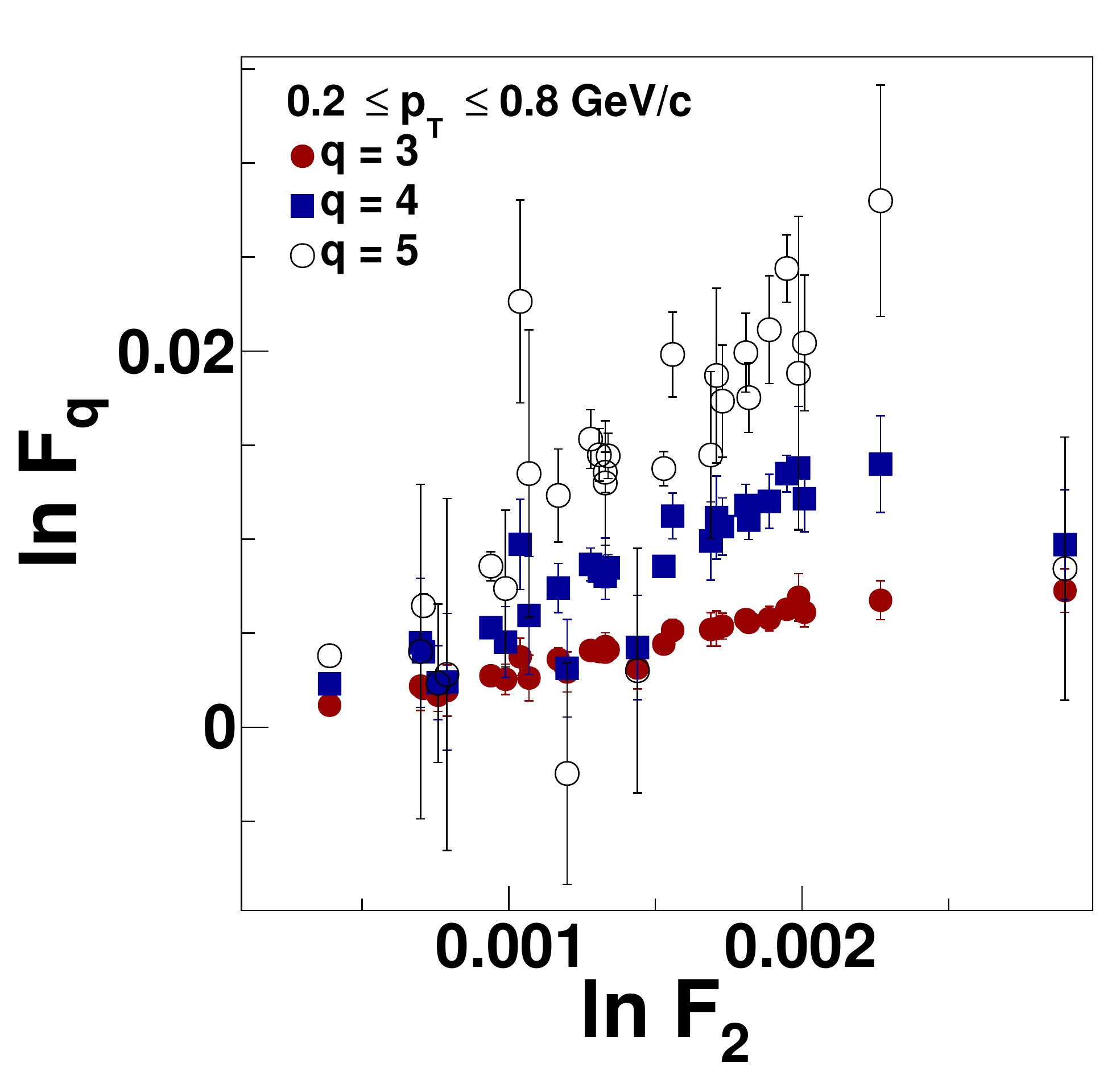}}
		\subfigure[\ 0.2 $\leq \pt \leq$ 1.0 \gev.]{
			\includegraphics*[scale=0.2]{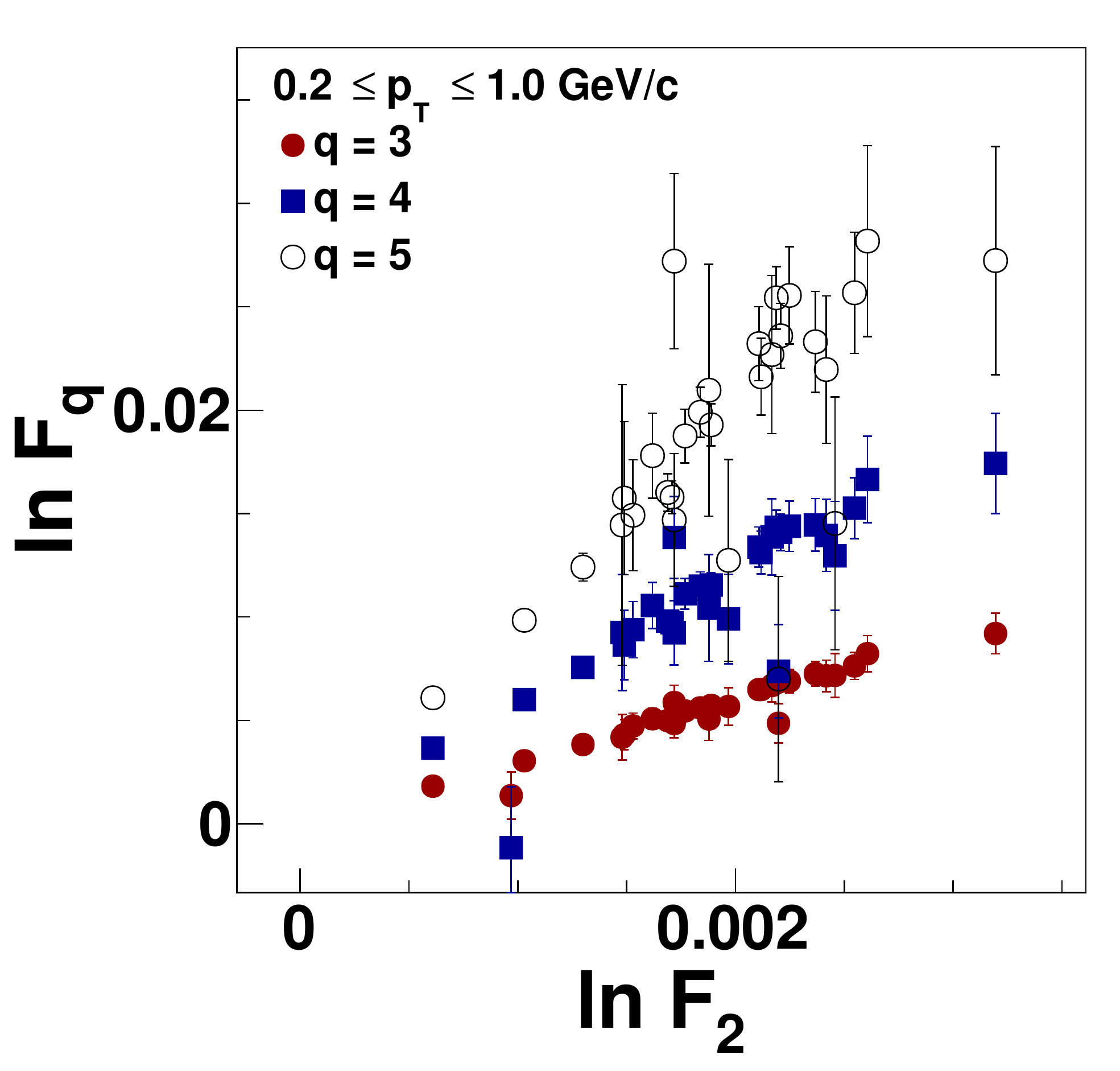}}
		\caption{\label{fig:fs/hydrocascade} Log-Log~\fq~dependence on \fqtwo~for EPOS3-hydro+cascade events for the \pt~bins 0.2 $\leq \pt \leq$ 0.8 \gev~and 0.2 $\leq \pt \leq$ 1.0 \gev.}
	\end{figure}

 
\fq~is observed to show a linear dependence on \fqtwo~even in the absence of M-scaling~\citep{Hwa:1992uq}. In Figures \ref{fig:fs/hydro} and \ref{fig:fs/hydrocascade}, $\ln F_{q}$ vs $\ln F_{2}$ plots are given for the bins with $\Delta \pt \geq 0.6$ \gev, the same bins in which for M-scaling is observed for low M values. $F_q$ is observed to follow power law in $F_2$ whereas, in the smaller \pt~bins, F-scaling is also absent.
	

Scaling index, $\nu$ is determined from the slope for $\ln \beta_q$ against $\ln (q-1)$. The scaling index, ($\nu$) obtained for the two cases are enlisted in Table (\ref{scalingindex}). The NA22 data on particle production in hadronic collisions gives $\nu=$1.45$\pm$0.04, heavy-ion experiments $\nu=$1.55$\pm$0.12~\citep{Hwa:1992uq} and $\nu=$1.459$\pm$0.021~\citep{Jain:1991mv}. 	
	However, the average value of $\nu$ obtained here is 1.795$\pm$0.156 EPOS3 (hydro) and 1.824$\pm$0.295 EPOS3 (hydro+cascade), which is different from the value 1.304 as is obtained from the GL formalism for the second order phase transition. The values obtained here are significant, since the lattice QCD predicts continuous crossover type of phase transition~\citep{Bhattacharya:2014ara}.
	\begin{table}[ht]
		\centering
		\caption{Scaling index values of the event samples.}
		\begin{tabular}{ccc}
			\hline
			Event Sample&$\pt$ bins(\gev)&Value of $\nu$\\
			\hline
			\hline
			\multirow{2}{*}{Hydro}&0.2-0.8&1.84$\pm$0.19\\
			&0.2-1.0&1.75$\pm$0.12\\
			\hline
			\multirow{2}{*}{Hydro+Cascade}&0.2-0.8&1.85$\pm$0.33\\
			&0.2-1.0&1.80$\pm$0.26\\
			\hline
		\end{tabular}
		\label{scalingindex}
	\end{table}
\par
\begin{figure}[h!]
		\centering
		
			\includegraphics[scale=0.45]{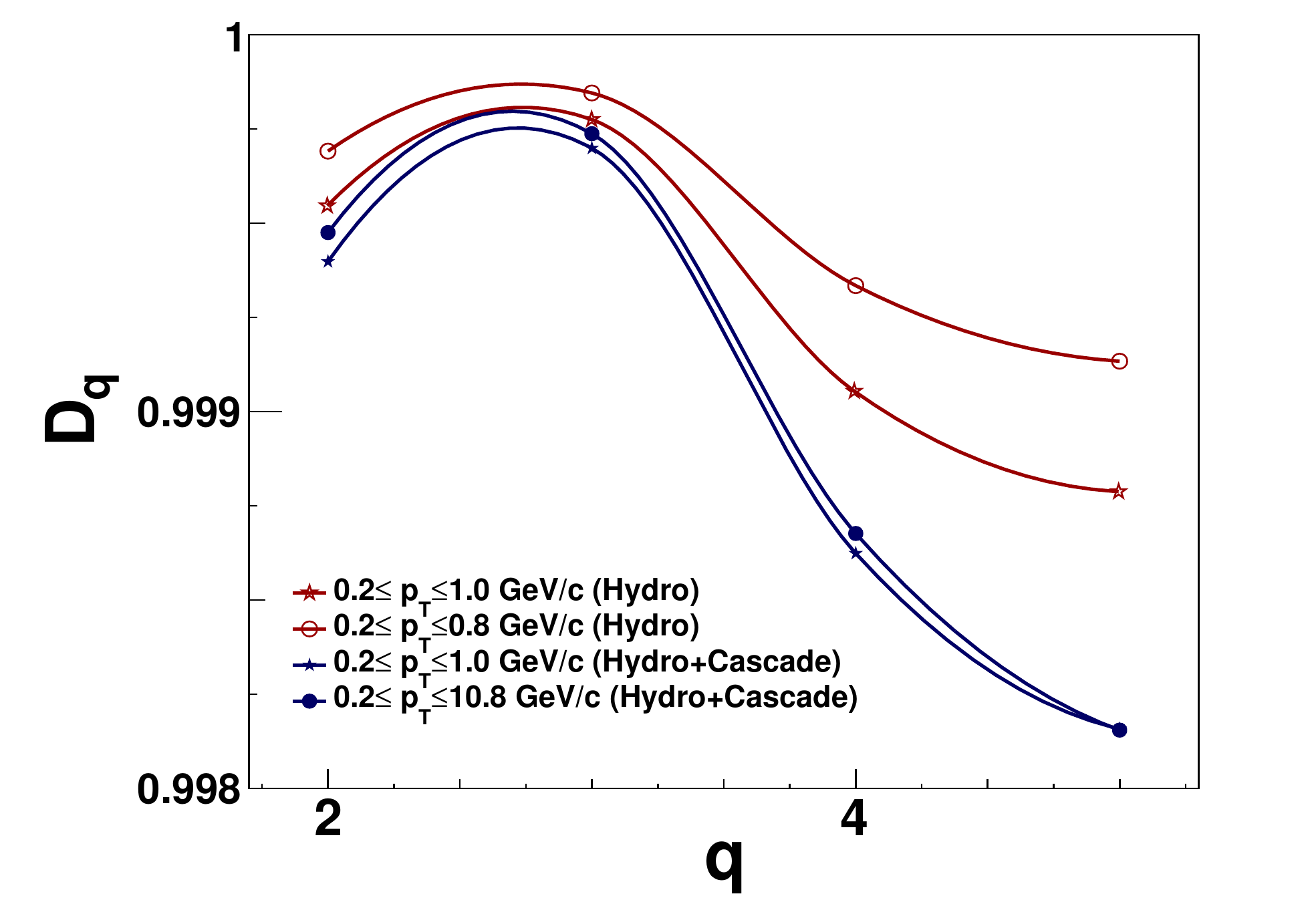}
		\caption{\label{fig:fractality} $q$ dependence of fractal dimensions, $D_q$ for the EPOS3 (hydro \& hydro+cascade) events in the two \pt~bins in which weak M-scaling and F-scaling is observed.}
	\end{figure}

 
For the two $\pt$ bins in which M-scaling is observed for the low M-region, the $d_q's$ have been calculated from the intermittency index ($\phi_{q}$) and thus the fractal dimensions $D_q$'s are determined  and are plotted against $q$ in Figure \ref{fig:fractality}. The $d_q$ grows in a way such that that the fractal (R$\grave{e}$nyi) dimensions $D_q$ are close to one. However in the data, the fractal dimensions are observed to be much smaller than one~\citep{Sarkisian:1993pi,Kittel:2005fu,DeWolf:1995nyp}. This observation indicates that EPOS3 in hydro and hydro+cascde mode do not have fractal behaviour.  The $D_q$ decreases faster with increasing order of the moment $q$ and has similar behaviour for both the bins for the two modes of the EPOS3 modes. $D_{2} < D_{q}$ for $q > 2$ further contradicts the data~\citep{Sarkisian:1993pi,Sarkisian:1993mf}.  

\section{Summary}
\label{summ}
 An event-by-event intermittency analysis is performed for the charged particle multiplicity distributions of the events generated using two different modes of EPOS3 hydrodynamical model. Central events with b~$\leq$~3.5 $fm$ generated from \PbPb~collisions at \sqrtSnn, have been studied. The two-dimensional intermittency analysis is performed in ($\eta, \phi$) phase space with $|\eta|\leq 0.8$ and full azimuth space in the narrow transverse momentum (\pt) bins in the region with \pt~$\leq 1.0$~\gev~with the objective to study the scaling behaviour of the charged particle multiplicity fluctuations as are introduced by the hydro and hydro+cascade modes of EPOS3 model. In narrow \pt~bins in the ($\eta, \phi$) space, M-scaling is found to be absent whereas a weak M-scaling in two larger \pt~bins with $\Delta \pt \geq 0.6$~\gev~viz; 0.2 $\leq$p$_T$$\leq 0.8$~\gev~and 0.2 $\leq$p$_T$$\leq$ 1.0 \gev, is observed. Absence of power-law of \fq~with $M$ indicates the absence of intermittency and hence self-similar behaviour in the local multiplicity fluctuations in charged particle generation in the events and hence the EPOS3 model. For the narrow \pt~bins $\Delta \pt < 0.6\, \gev$, F-scaling which is independent of the observation of M-scaling, is also absent. However, in the wider \pt~ bins 0.2 $\leq \pt \leq$ 0.8~\gev~and 0.2 $\leq \pt \leq$ 1.0~\gev, \fq~shows power law with~\fqtwo. This is in contrast to what is observed in \cite{Sharma:2018vtf}, where M-scaling as well as F-scaling is observed in the small $p_{T}$ bins with $\Delta p_{T} \le 0.2$~\gev. The average value of $\nu$, the scaling exponent for these two bins from the two modes of EPOS3 is 1.809, a value different from 1.304, the value as obtained from Ginzburg-Landau theory for second-order phase transition. This suggests absence of spatial fluctuations in the local charged particle generation, that was not the case with the transport String Melting AMPT model~\cite{Sharma:2018vtf}. In the larger phase space bins corresponded to $\Delta \pt \geq 0.6$~\gev~in the low \pt~region M-scaling obseved for the low $M$ values is reflected in the value of fractal dimension, $D_q$. $D_q$ shows an inverse dependence on $q$ for $q > 2$, thus the presence of multifractality in the larger phase space bins. This is in contrast to the observations at lower energies. Similar studies of experimental data from RHIC and LHC are yet not available. It would be interesting to see whether we get similar observations from the experiment or not.
\section{Data Availability}
The data used to support the findings of this study are available from the corresponding author upon request.
\section{Acknowledgements} 
\label{ackno}
One of the authors is thankful to Prof. Rudolph C. Hwa for his discussions on the analysis. The authors thankfully acknowledge Tanguy Peirog, Klaus Werner, Yuri Karpenko for their assistance in EPOS3 installation. Our sincere thanks to the Grid computing facility at VECC-Kolkata, India to facilitate the generation of the Monte Carlo events for this work.
	\onecolumngrid{

	}

\begin{thebibliography}{99}
			\bibitem{Sarkar:2010zza} 
			S.~Sarkar, H.~Satz and B.~Sinha,
			``The physics of the quark-gluon plasma,''
			Lect.\ Notes Phys.\  {\bf 785}, pp.1 (2010).
			doi:10.1007/978-3-642-02286-9

			\bibitem{Burnett:1983pb} 
			T.~H.~Burnett {\it et al.},
			``Extremely High Multiplicities in High-Energy Nucleus Nucleus Collisions,''
			Phys.\ Rev.\ Lett.\  {\bf 50}, 2062 (1983).
			doi:10.1103/PhysRevLett.50.2062

			\bibitem{Kittel:2005fu} 
			W.~Kittel and E.~A.~De Wolf,
			``Soft multihadron dynamics,''
			World Scientific (Singapore, 2005).

			\bibitem{Bialas:1985jb} 
			A.~Bialas and R.~B.~Peschanski,
			``Moments of Rapidity Distributions as a Measure of Short Range Fluctuations in High-Energy Collisions,''
			Nucl.\ Phys.\ B {\bf 273}, 703 (1986).
			doi:10.1016/0550-3213(86)90386-X
			
			\bibitem{Bialas:1988wc} 
			A.~Bialas and R.~B.~Peschanski,
			``Intermittency in Multiparticle Production at High-Energy,''
			Nucl.\ Phys.\ B {\bf 308}, 857 (1988).
			doi:10.1016/0550-3213(88)90131-9
			
			\bibitem{fractals} J.~Brickmann,``B. Mandelbrot: The Fractal Geometry of Nature, Freeman and Co., San Francisco 1982. 460 Seiten,''
			Berichte der Bunsengesellschaft für physikalische Chemie {\bf 89}, 2 (1985)
			doi:10.1002/bbpc.19850890223
			[https://onlinelibrary.wiley.com/doi/abs/10.1002/bbpc.19850890223].
			\bibitem{Lindaereichl} 
			L.~Reichl,
			``A modern course in statistical physics,''
			American Journal of Physics {\bf 67}, 12 (1999).
			\bibitem{Satz:1989vj} 
			H.~Satz,
			``Intermittency and Critical Behavior,''
			Nucl.\ Phys.\ B {\bf 326}, 613 (1989).
			doi:10.1016/0550-3213(89)90546-4
			\bibitem{Bambah:1989fy} 
			B.~Bambah, J.~Fingberg and H.~Satz,
			``The Onset of Intermittent Behavior in the Ising Model,''
			Nucl.\ Phys.\ B {\bf 332}, 629 (1990).
			doi:10.1016/0550-3213(90)90004-W
			\bibitem{Bialas:1990xd} 
			A.~Bialas and R.~C.~Hwa,
			``Intermittency parameters as a possible signal for quark - gluon plasma formation,''
			Phys.\ Lett.\ B {\bf 253}, 436 (1991).
			doi:10.1016/0370-2693(91)91747-J
			\bibitem{DeWolf:1995nyp} 
			E.~A.~De Wolf, I.~M.~Dremin and W.~Kittel,
			``Scaling laws for density correlations and fluctuations in multiparticle dynamics,''
			Phys.\ Rept.\  {\bf 270}, 1 (1996)
			doi:10.1016/0370-1573(95)00069-0
			[hep-ph/9508325].
			
			\bibitem{Deppman:2019klo} 
			A.~Deppman, E.~Megías and D.~P.~Menezes,
			``Fractal structure in Yang-Mills fields and non extensivity,''
			arXiv:1905.06382 [hep-th].
			\bibitem{Hwa:1992uq} 
			R.~C.~Hwa and M.~T.~Nazirov,
			``Intermittency in second order phase transition,''
			Phys.\ Rev.\ Lett.\  {\bf 69}, 741 (1992).
			doi:10.1103/PhysRevLett.69.741
			\bibitem{Werner:2013tya} 
			K.~Werner, B.~Guiot, I.~Karpenko and T.~Pierog,
			``Analysing radial flow features in p-Pb and p-p collisions at several TeV by studying identified particle production in EPOS3,''
			Phys.\ Rev.\ C {\bf 89}, no. 6, 064903 (2014)
			doi:10.1103/PhysRevC.89.064903
			[arXiv:1312.1233 [nucl-th]].
			
			\bibitem{Werner:2010aa} 
			K.~Werner, I.~Karpenko, T.~Pierog, M.~Bleicher and K.~Mikhailov,
			``Event-by-Event Simulation of the Three-Dimensional Hydrodynamic Evolution from Flux Tube Initial Conditions in Ultrarelativistic Heavy Ion Collisions,''
			Phys.\ Rev.\ C {\bf 82}, 044904 (2010)
			doi:10.1103/PhysRevC.82.044904
			[arXiv:1004.0805 [nucl-th]].
			
			\bibitem{Werner:2007bf} 
			K.~Werner,
			``Core-corona separation in ultra-relativistic heavy ion collisions,''
			Phys.\ Rev.\ Lett.\  {\bf 98}, 152301 (2007)
			doi:10.1103/PhysRevLett.98.152301
			[arXiv:0704.1270 [nucl-th]].
			\bibitem{Drescher:2000ha} H.~J.~Drescher, M.~Hladik, S.~Ostapchenko, T.~Pierog and K.~Werner,``Parton based Gribov-Regge theory,''
			Phys.\ Rept.\  {\bf 350}, 93 (2001)
			doi:10.1016/S0370-1573(00)00122-8
			[hep-ph/0007198].
			\bibitem{Cooper:1974mv} 
			F.~Cooper and G.~Frye,
			``Comment on the Single Particle Distribution in the Hydrodynamic and Statistical Thermodynamic Models of Multiparticle Production,''
			Phys.\ Rev.\ D {\bf 10}, 186 (1974).
			doi:10.1103/PhysRevD.10.186
		
			\bibitem{eneziano} 
			G~Veneziano,
			``Proc. 3rd Workshop on Current Problems in High Energy Particle
			Theory, Florence,''
			Eds.\ R.\ Casalbuoni\ et\ al.,\ Johns\ Hopkins\ University\ Press\ {\bf 45}, (1979).

			\bibitem{ATLAS:2011ag} 
			G.~Aad {\it et al.} [ATLAS Collaboration],
			``Measurement of the centrality dependence of the charged particle pseudorapidity distribution in lead-lead collisions at $\sqrt{s_{NN}}=2.76$ TeV with the ATLAS detector,''
			Phys.\ Lett.\ B {\bf 710}, 363 (2012)
			doi:10.1016/j.physletb.2012.02.045
			[arXiv:1108.6027 [hep-ex]].
			
		
			
			\bibitem{Hwa:2014jea} 
			R.~C.~Hwa,
			``Recognizing Critical Behavior amidst Minijets at the Large Hadron Collider,''
			Adv.\ High Energy Phys.\  {\bf 2015}, 526908 (2015)
			doi:10.1155/2015/526908
			[arXiv:1411.6083 [nucl-ex]].
			
			\bibitem{Hwa:2016khr} 
			R.~C.~Hwa and C.~B.~Yang,
			``Observable Properties of Quark-Hadron Phase Transition at the Large Hadron Collider,''
			Acta Phys.\ Polon.\ B {\bf 48}, 23 (2017)
			doi:10.5506/APhysPolB.48.23
			[arXiv:1601.04671 [nucl-th]].
			
			
			
			\bibitem{Hwa:2011bu} 
			R.~C.~Hwa and C.~B.~Yang,
			``Local Multiplicity Fluctuations as a Signature of Critical Hadronization at LHC,''
			Phys.\ Rev.\ C {\bf 85}, 044914 (2012)
			doi:10.1103/PhysRevC.85.049904, 10.1103/PhysRevC.85.044914
			[arXiv:1111.6651 [nucl-th]].
			
			
			
			\bibitem{PALADIN1987147} 
			G.~Paladin and A.~Vulpiani,
			``Anomalous scaling laws in multifractal objects,''
			Physics\ Reports\  {\bf 156}, 147-225 (1987).
			doi:10.1016/0370-1573(87)90110-4
		
			\bibitem{Feder1988} 
			J.~Feder,
			``The Fractal Dimension,''
			Springer\ US,\ Boston,\ MA, 6-30 (1988).
			\bibitem{Chiu:1990bc} 
			C.~B.~Chiu, K.~Fialkowski and R.~C.~Hwa,
			``Nonstatistical component of the multifractal spectral function,''
			Mod.\ Phys.\ Lett.\ A {\bf 5}, 2651 (1990).
			doi:10.1142/S0217732390003085
			
			
			
			\bibitem{Sharma:2018vtf} 
			R.~Sharma and R.~Gupta,
			``Scaling Properties of Multiplicity Fluctuations in the AMPT Model,''
			Adv.\ High Energy Phys.\  {\bf 2018}, 6283801 (2018)
			doi:10.1155/2018/6283801
			[arXiv:1806.10854 [hep-ph]].
				\bibitem{Metzger}
			W.J.~Metzger
			"Estimating the Uncertainties of Factorial Moments,"
			preprint HEN-455 (Nijmegen University, 2004).\
			\bibitem{Jain:1991mv} 
			P.~L.~Jain and G.~Singh,
			``One-dimensional and two-dimensional analysis of intermittency in ultrarelativistic nucleus-nucleus interactions,''
			Phys.\ Rev.\ C {\bf 44}, 854 (1991).
			doi:10.1103/PhysRevC.44.854
			
			\bibitem{Bhattacharya:2014ara} 
			T.~Bhattacharya {\it et al.},
			``QCD Phase Transition with Chiral Quarks and Physical Quark Masses,''
			Phys.\ Rev.\ Lett.\  {\bf 113}, no. 8, 082001 (2014)
			doi:10.1103/PhysRevLett.113.082001
			
			\bibitem{Sarkisian:1993pi} 
			E.~K.~Sarkisyan, L.~K.~Gelovani and G.~G.~Taran,
			``Fractality and fluctuations in charged particle pseudorapidity distributions in central C (Ne, Cu) collisions at 4.5-A/GeV/c,''
			Phys.\ Lett.\ B {\bf 302}, 331 (1993).
			doi:10.1016/0370-2693(93)90404-6
			
			\bibitem{Sarkisian:1993mf} 
			E.~K.~Sarkisyan, L.~K.~Gelovani, G.~I.~Sakharov and G.~G.~Taran,
			``Fractal analysis of pseudorapidity fluctuations in 4.5-A/GeV/c C (Ne, Cu) central collisions,''
			Phys.\ Lett.\ B {\bf 318}, 568 (1993).
			doi:10.1016/0370-2693(93)91557-4
			
		\end{thebibliography}
\end{document}